Phantom-Chain Simulations for the Effect of Node Functionality on the Fracture of Star-Polymer Networks

[1*]Yuichi Masubuchi, [1]Yuya Doi, [1]Takato Ishida, [2]Naoyuki Sakumichi, [2]Takamasa Sakai, [3]Koichi Mayumi, [4]Kotaro Sato, and [1]Takashi Uneyama

[1]Department of Materials Physics, Nagoya University, Nagoya 4648603, JAPAN

[2]Department of Chemistry and Biotechnology, The University of Tokyo, Tokyo 1138654, JAPAN

[3]The Institute for Solid State Physics, The University of Tokyo, Chiba, 2778581 Japan

[4]Department of Chemical Science and Engineering, Tokyo Institute of Technology, Tokyo 1528550, Japan

*to whom correspondence should be addressed
mas@mp.pse.nagoya-u.ac.jp
TEL: +81-527892551

**ABSTRACT**

The influence of node functionality ($f$) on the fracture of polymer networks remains unclear. While many studies have focused on multi-functional nodes with $f > 4$, recent research suggests that networks with $f = 3$ exhibit superior fracture properties compared to those with $f = 4$. To clarify this discrepancy, we conducted phantom chain simulations for star-polymer networks varying $f$ between 3 and 8. Our simulations utilized equimolar binary mixtures of star branch prepolymers with a uniform arm length. We employed a Brownian dynamics scheme to equilibrate sols and induce gelation through end-linking reactions. We prevented the formation of odd-order loops owing to the binary reaction and second-order loops algorithmically. We stored network structures at various conversion ratios ($\varphi_c$) and minimized energy to reduce computation costs induced by structural relaxation. We subjected the networks to stretching until fracture to determine stress and strain at break and work for fracture, $\varepsilon_b$, $\sigma_b$, and $W_b$. These fracture characteristics are highly dependent on $\varphi_c$ for networks with small $f$ but relatively insensitive for those with large $f$. Thus, the networks with small $f$ exhibit greater fracture properties than those with large $f$ at high $\varphi_c$, whereas the opposite relationship occurs at low $\varphi_c$. We analyzed $\varepsilon_b$, $\sigma_b$, and $W_b$ concerning the cycle rank $\xi$ and the broken strand fraction $\varphi_{bb}$. We found $\varepsilon_b$, $\sigma_b/\varphi_{bb}$, and $W_b/\varphi_{bb}$ monotonically decrease with increasing $\xi$, and the data for various $f$ and $\varphi_c$ superpose with each other to draw master curves. These results imply that the mechanical superiority of the networks with small $f$ comes from their smaller $\xi$ that gives higher $\varepsilon_b$, $\sigma_b/\varphi_{bb}$, and $W_b/\varphi_{bb}$ than the networks with large $f$.

**INTRODUCTION**

Although the effect of node functionality on the modulus of polymer networks has been well established[1–4], its impact on network fracture remains unclear. Numerous attempts have been made to develop tough network polymers with multiple node functionalities[5], which are attained by hyper-branched polymers[6], copolymer micelles[7], polymer crystals[8], nano crays[9], and microgels[10]. These materials exhibit superior mechanical properties compared to randomly cross-linked ones [5]. Meanwhile, a recent report by Fujiyabu et al.[11] challenges this widely-employed direction by showing that hydrogels made from star-branched prepolymers with node functionality $f = 3$ realize better toughness than the $f = 4$ analogs. Although they attribute this result to stretch-induced crystallization observed only for $f = 3$, the mechanism has yet to be fully elucidated.

There have been some theoretical attempts for network fracture. Kuhn's theory[12] provides a framework based on the strand length before stretching. In an ideal network where all strands connecting network nodes have equal lengths, strain at break can be calculated as the ratio of the maximum length for the break to the length before stretching. The effect of node functionality on network fracture is reflected in the difference in the distribution of strand lengths before the stretch. Coarse-grained molecular dynamics simulations performed by Tsige and Stevens[13,14] have demonstrated that networks made from mixtures of linear prepolymers and star-branched cross-linkers with $f = 3$ exhibit larger fracture energy than those with $f = 4$ and 6. This superiority of $f = 3$ is attributed to the shorter distance between network nodes before the stretch. Fujiyabu et al.[11] discussed this direction to explain their results partly. Another critical factor is the number of defects in the network. Lin and Zhao[15] proposed a theory for imperfect networks containing defects such as dangling ends and loops. In their view, fracture energy increases with the number of secondary loops. (A secondary loop corresponds to connectivity between network nodes that share two strands.) This prediction is consistent with the theory by Barney et al.[16], who extended the real elastic network theory (RENT) proposed by Zhong et al.[17] Since the number of secondary loops increases with the number of arms for star-branch prepolymers, their theory supports the direction of multiple node functionalities while contradicting the result reported by Fujiyabu et al. [11]

Motivated by the work of Fujiyabu et al. [11], we previously performed phantom chain simulations where networks with $f = 3$ and 4 were constructed via the gelation of star-branched prepolymers[18]. The simulation of gelation was made with a Brownian dynamics scheme. Afterward, we imposed energy minimization to the obtained networks under stretch to skip structural relaxation induced by every breakage of strands. The results revealed that the fracture energy of $f = 3$ networks is larger than that of $f = 4$ analogs, even without crystallization. In this study, the superiority of $f = 3$ to $f = 4$ was attributed to the difference in the strand length distribution, following Kuhn's abovementioned argument. However, the difference was faint, and the physics behind this phenomenon has yet to be fully clarified.

In this study, we extended our previous work to larger $f$ values to systematically investigate the effect of $f$ on network fracture. We performed phantom chain simulations for networks composed of star-branch prepolymers by changing $f$ from 3 to 8. We stored network structures with various conversion rates $\varphi_c$ for each $f$ during the gelation. We imposed energy minimization and stretched to the obtained networks until the break. We measured strain and stress at the break, $\varepsilon_b$ and $\sigma_b$, and acquired fracture energy $W_b$ by numerically integrating the stress-strain curve. The results revealed that the relation between the characteristic quantities for fracture and node functionality strongly depends on the conversion ratio, to which the inconsistency among earlier studies can be attributed. We further analyzed the data concerning the fraction of broken bonds $\varphi_{bb}$ and the cycle rank $\xi$ to report that $\varepsilon_b$, $\sigma_b/\varphi_{bb}$, and $W_b/\varphi_{bb}$ can be superposed to master curves as functions of $\xi$. Details are shown below.

**MODEL AND SIMULATOINS**
We consider blends of star-branch prepolymers that emerge end-linking reactions only between different chemistries. Binary star polymers as the precursor have been widely employed to eliminate primary loop formation[19,20]. Although not frequently stated, higher odd-order loops (like 3rd and 5th-order loops)[17,21] are also eliminated owing to the binary nature. Besides, in this specific study, we prohibited the formation of 2nd-order loops (secondary loops) to eliminate their effect[15]. Because earlier experiments[22,23] have demonstrated that the number of trapped entanglements is negligible with a tuned molecular weight of the branching arm, we employ phantom chains, neglecting excluded volume interactions.

We distributed $M$ prepolymers with $f$ branching arms, each consisting of $N_a$ beads in the simulation box with periodic boundary conditions. According to the previous study[18], we chose $N_a = 5$ and $M = 1600$, whereas we varied $f$ ranging from 3 to 8. The effects of system size and prepolymer molecular weight were discussed in our previous study [18]. The simulation box dimension was set to realize the bead number density $\rho$, chosen at 8 unless stated. This bead density corresponds to $c/c^* \sim 4$. The prepolymer sols were sufficiently equilibrated by a Brownian dynamics scheme, in which the equation of motion for the bead position $\mathbf{R}_i$ was as follows.

$$\mathbf{0} = -\zeta \dot{\mathbf{R}}_i + \frac{3k_B T}{a^2} \sum_k f_{ik} \mathbf{b}_{ik} + \mathbf{F}_i \qquad (1)$$

Here, $\zeta$ is the friction coefficient, $a$ is the average bond length, and $\mathbf{b}_{ik} \equiv \mathbf{R}_i - \mathbf{R}_k$ is the bond vector between connected beads. $f_{ik}$ is the non-linear spring factor for finite chain extensibility written as $f_{ik} = (1 - \mathbf{b}_{ik}^2/b_{\max}^2)^{-1}$. We employed this non-linear spring to avoid bond elongation beyond the critical length for scission introduced later. Following the previous study [18], we set $b_{\max} = 2$. $\mathbf{F}_i$ is Gaussian random force obeying $\langle \mathbf{F}_i \rangle = \mathbf{0}$ and $\langle \mathbf{F}_i(t) \mathbf{F}_j(t') \rangle = 2 k_B T \delta_{ij} \delta(t-t') \mathbf{I}/\zeta$. We chose $a$, $k_B T$, and $\tau = \zeta a^2/k_B T$ as units of length, energy, and time. We employed a 2nd-order scheme[24] for the numerical integration with the step size $\Delta t = 0.01$.

After equilibration, we initiated end-linking reactions, in which end beads were connected when a pair of end beads came closer to each other. For this reaction, we chose the critical distance $r_c = 0.5$, and the reaction rate $p = 0.1$, according to previous studies[18,25,26]. As mentioned, this reaction did not occur between molecules with the same chemistry; thus, primary and higher odd-order loops were eliminated. In addition, we prohibited 2nd-order loop formation algorithmically. Namely, when a pair of arms were already connected for the subjected prepolymers, we disallowed the reaction between the other arm pairs. The reaction simulations were performed until the conversion rate $\varphi_c$ reached 0.95. During the simulation, the network structures with various $\varphi_c$ between 0.6 and 0.95 were stored. Note that we had the same bond potential, including finite extensibility, for the newly created bonds by the reaction as well as the originally existing ones.

We uniaxially elongated the obtained networks until the break. Following earlier studies[18,27–30], we switched off the reaction and the Brownian motion and minimized the total potential energy.

This strategy allows us to reduce computation costs induced by slow structural relaxations after each bond breakage, although we cannot discuss the dynamics and energy dissipation. The potential energy is given below.

$$U = -\frac{3k_B T b_{max}^2}{2a^2} \sum_{i,k} \ln\left(1 - \frac{\mathbf{b}_{ik}^2}{b_{max}^2}\right) \qquad (2)$$

This energy is consistent with the bond tension in eq 1 (including $f_{ik}$), and energy minimization corresponds to the minimization of total bond length. We employed the Broyden-Fletcher-Goldfarb-Sanno (BFGS) method[31] with the energy conversion parameter and the bead displacement parameter chosen at $\Delta u = 10^{-4}$ and $\Delta r = 10^{-2}$. See our previous study[18] for the choice of these parameter values. We alternatively applied stepwise infinitesimal affine elongation and imposed energy minimization. After each energy minimization process was finished, we observed the bond length. When the length of a bond exceeded the critical value $b_c = \sqrt{1.5}$, we removed the bond. With this $b_c$ value, we did not observe any bond scission in all the obtained networks before imposing elongation, owing to the finite chain extensibility mentioned above. When we detected bond scissions, we repeated the minimization step without deformation.

**RESULTS**

Figure 1 shows a typical example of the performed simulation. Panel (a) exhibits the prepolymers with 8 branching arms ($f = 8$) and two different chemistries, colored in red and blue. For clarity, the branching arms are displayed in a stretched state. These prepolymers were mixed in a simulation box with 800 molecules for each chemistry. For the bead number density $\rho = 8$, the box dimension is ca. $20^3$, sufficiently larger than the prepolymers under equilibrium. We equilibrated the system with the Brownian dynamics scheme and initiated gelation. The gelation process was also traced with the Brownian dynamics until the conversion rate reached $\varphi_c = 0.9$. The formation of loops up to 3rd-order was prohibited during the gelation. The resultant gel is shown in panel (b), where the system is reasonably homogeneous. We imposed energy minimization on the obtained network; the result is shown in panel (c). Then we applied a stepwise small elongational strain and energy minimization alternatively to the system, as shown in panels (d) and (e), which exhibit the network at $\varepsilon = 0.5$ and 1.0, respectively. Panels (h) and (i) show the development of stress $\sigma$ and unconnected strand fraction $\varphi_u$ during the stretch. We employ true stress and true strain to be consistent with our previous work[18]. Note that $\varphi_u$ starts at 0.1, consistent with $\phi_c = 0.9$. During the stretch, we monitored the bond length to remove bonds for which the length exceeded $b_c$. Because each bond elimination induces structural change, we repeated the energy minimization step without stretch while we found bond scission. Panel (f) shows the snapshot at $\varepsilon = 1.5$, where $\varphi_u \sim 0.11$, as seen in panel (h). The network was finally depercolated at $\varepsilon = 1.64$ with $\varphi_u = 1.58$ for this specific case. Panel (g) exhibited the structure just after eliminating the last bonds. Some remaining antler-like structures withdrew in the following energy minimization (snapshot not shown).

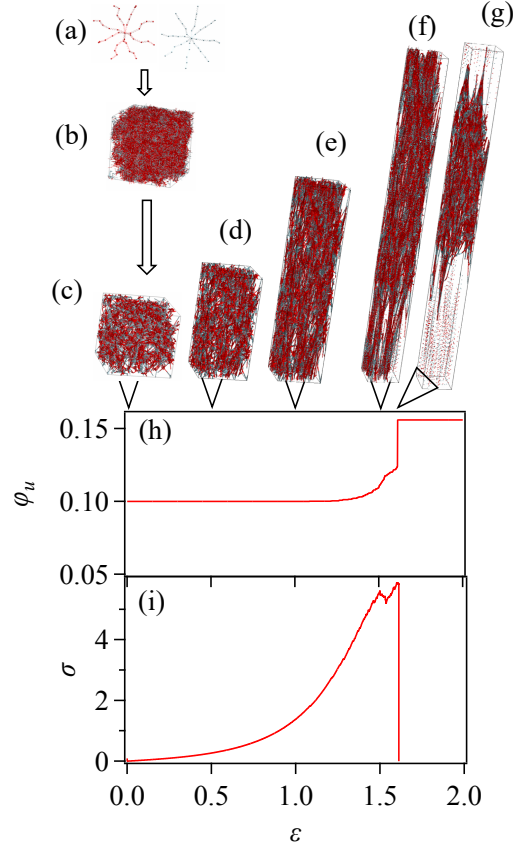

**Figure 1** Typical snapshots of prepolymers (a), the gelated network (b), the energy-minimized structure (c), the stretched states (d)-(f), the broken network (g), and development of unconnected strand fraction $\varphi_u$ (h) and stress $\sigma$ (i) during the stretch plotted against strain $\varepsilon$ for the case with $f = 8$ and $\varphi_c = 0.9$.

The stress-strain relation is strongly dependent on $f$ and $\varphi_c$. Figure 2 shows the cases for $f = 3$ (red) and 8 (black) with $\varphi_c = 0.9$ (solid) and 0.6 (broken). Each curve is an individual simulation run. With $\varphi_c = 0.9$, $f = 3$ attains larger strain and stress at break ($\varepsilon_b$ and $\sigma_b$) than $f = 8$. The work for fracture $W_b$, obtained by numerically integrating the curve until the break, is larger for $f = 3$ than $f = 8$. With decreasing $\varphi_c$, modulus decreases, and $\varepsilon_b$ increases for both cases, but the change is more significant for $f = 3$ than $f = 8$. Consequently, with $\varphi_c = 0.6$, $\sigma_b$ and $W_b$ for $f = 8$ are larger than those for $f = 3$.

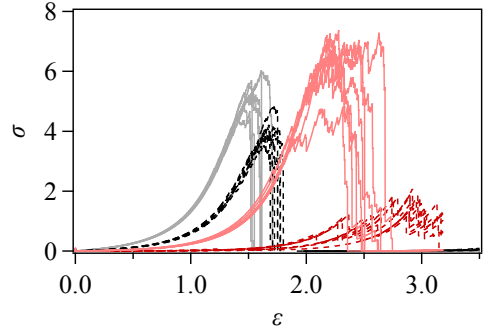

**Figure 2** Stress-strain curves for $f = 3$ (red) and 8 (black) with $\varphi_c = 0.9$ (solid) and 0.6 (broken). Each curve corresponds to the individual simulation run.

Figure 3 shows $\varepsilon_b$, $\sigma_b$, and $W_b$ plotted against $\varphi_c$ for $f = 3$ (red circle) and 8 (black circle). As mentioned in Fig 2, with increasing $\varphi_c$, $\varepsilon_b$ decreases, and $\sigma_b$ and $W_b$ increase. Overall, $f = 8$ networks are robust as the obtained fracture characteristics are not significantly dependent on $\varphi_c$. The robustness is because the network is well-developed to sustain stress even at low $\varphi_c$ owing to the high functionality. This result is in harmony with earlier studies[5] that claim the advantages of multi-functional network nodes. In contrast, the mechanical properties for $f = 3$ networks are sensitive to $\varphi_c$. When $\varphi_c$ is small, the network structure does not sufficiently develop, and a small fraction of strands realize the percolation. The work for fracture $W_b$ to break such strands is relatively small. $\varepsilon_b$ becomes large because of prepolymers for which only two arms are connected to the others, as discussed later. As $\varphi_c$ increases, the number of active strands significantly increases than the case with $f = 8$. Due to this network development, $\varepsilon_b$ decreases, and $\sigma_b$ and $W_b$ increase steeper than $f = 8$. Consequently, with $\varphi_c \geq 0.7$, $\sigma_b$ and $W_b$ for $f = 3$ are larger than for $f = 3$.

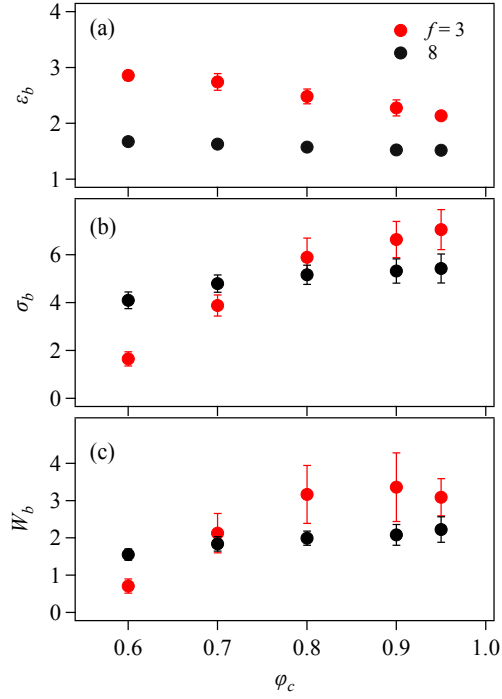

**Figure 3** Strain at break $\varepsilon_b$ (a), stress at break $\sigma_b$ (b), and work for fracture $W_b$ (c) as functions of conversion rate $\varphi_c$ for $f = 3$ (red circle) and 8 (black circle). Error bars exhibit standard deviation among eight independent simulation runs.

Figure 4 shows $\varepsilon_b$, $\sigma_b$, and $W_b$ as functions of $f$ with $\varphi_c = 0.6$ (filled triangle) and 0.9 (unfilled triangle). $\varepsilon_b$ systematically decreases with increasing $f$, whereas $\sigma_b$ and $F_b$ exhibit non-monotonic behavior against $f$, particularly for $\varphi_c = 0.6$ (see filled triangle). As mentioned in Fig 3, $f = 3$ networks are premature at $\varphi_c = 0.6$, resulting in small $\sigma_b$ and $W_b$ values. As $f$ increases, the effective network grows to sustain stress, and thus, $\sigma_b$ and $W_b$ increase in the range of $3 \leq f \leq 6$. However, for higher $f$ values, the developed network depresses the stretchability; thus, $W_b$ decreases with increasing $f$. For the case with $\varphi_c = 0.9$ (see unfilled triangle), the network develops even for $f = 3$. At this conversion ratio, $\varepsilon_b$, $\sigma_b$, and $W_b$ monotonically decrease with increasing $f$.

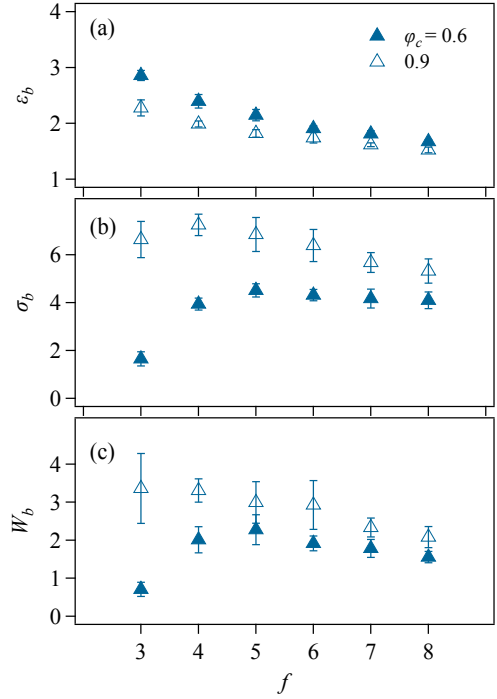

**Figure 4** Strain at break $\varepsilon_b$ (a), stress at break $\sigma_b$ (b), and work for fracture $W_b$ (c) as functions of prepolymer functionality $f$ for $\varphi_c = 0.6$ (filled triangle) and 0.9 (unfilled triangle). Error bars exhibit standard deviation among eight independent simulation runs.

**DISCUSSION**

The primary outcome is that networks with small $f$ appear mechanically superior to those with large $f$ when $\varphi_c$ is large. As qualitatively mentioned above, the fracture behavior is related to the network maturity, and we attempt to characterize it by the cycle rank $\xi$ per prepolymer in the percolated network. We determined $\xi$ from the number of effective strands $\nu$ and effective nodes $\mu$ (per prepolymer) as $\xi = \nu - \mu$, following the earlier study[32,33]. We included prepolymers with two reacted arms for further analysis, even though they are excluded according to the Scanlan-Case criterion[34,35]. Nonetheless, inclusion does not affect the value of $\xi$ because for such prepolymers $\nu$ and $\mu$ are the same.

Figure 5 shows $\xi$ for our examined networks by symbols. Solid curves represent the theoretical prediction according to the mean-field theory[33,36,37], with which our $\xi$ values are fully consistent. Based on the results, our networks align statistically with the theoretical setup; we do not include odd-ordered and second-ordered loops, which are also disregarded in the mean-field theory. Additionally, the higher-even-ordered loops seem to have minimal impacts. Appendix A shows the modulus $G$ concerning $\xi$.

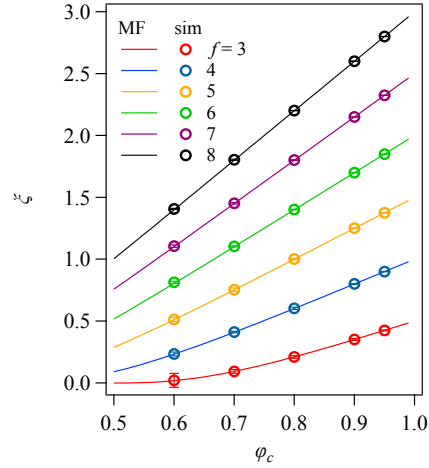

**Figure 5** Cycle rank $\xi$ plotted against conversion ratio $\varphi_c$ for various node functionality $f$. Symbols are the results obtained from the simulated networks. Solid curves are the predictions of the mean-field theory[33]. Error bars (mostly within the symbols) show the standard deviations among the eight independent simulation runs.

Figure 6 shows the fracture characteristics plotted against $\xi$. Panel (a) exhibits that strain at break $\varepsilon_b$ for various $f$ and $\varphi_c$ can be superposed to a master curve if those are plotted against $\xi$. Although we have no theoretical explanation, the relationship between $\varepsilon_b$ and $\xi$ can be apparently fitted by a power-law function $\varepsilon_b = 1.85\xi^{-0.2}$ drawn by the red solid curve. Nevertheless, this result implies that the network maturity is characterized by $\xi$ for stretchability.

Figures 6 (b) and (c) exhibit stress at break $\sigma_b$ and work for fracture $W_b$ plotted against $\xi$. $\sigma_b$ and $W_b$ increases with increasing $\xi$ for each $f$. As $f$ increases, the increment of $\sigma_b$ and $W_b$ against $\xi$ decreases systematically. According to the Lake-Thomas theory[38], these characteristics of network fracture can be related to bond breakage. Indeed, the ratio of broken bonds at the fracture $\varphi_{bb}$ exhibits similar behavior with $\sigma_b$ and $W_b$ concerning $\xi$ and $f$, as shown in Fig 6 (d).

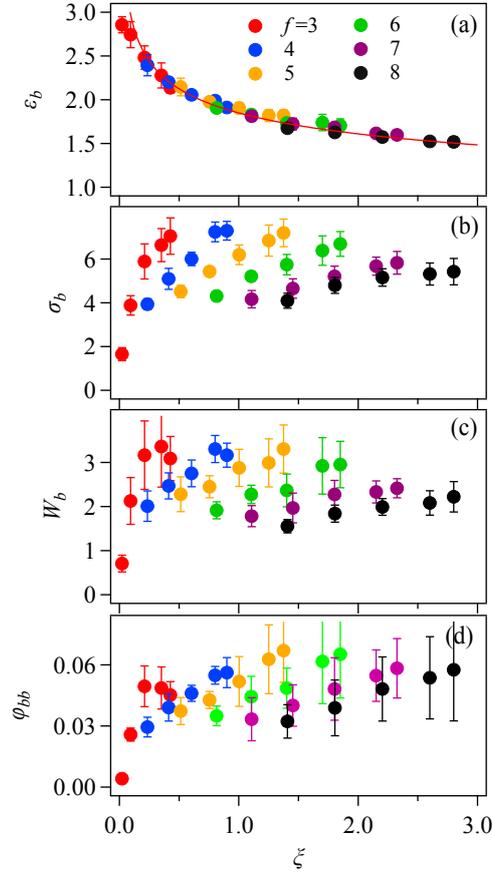

**Figure 6** Strain at break $\varepsilon_b$ (a), stress at break $\sigma_b$ (b), work for fracture $W_b$ (c), and broken bond fraction $\varphi_{bb}$ (d) obtained for various $f$ and $\varphi_c$ plotted against the cycle rank. Error bars exhibit the standard deviations for eight independent simulation runs. The red solid curve in panel (a) displays an apparent relation written as $\varepsilon_b = 1.85\xi^{-0.2}$.

Seeing Figs 6 (b)-(d), we found that master curves can be drawn for $\sigma_b$ and $W_b$ if these quantities are normalized by $\varphi_{bb}$, as shown in Fig 7. These panels demonstrate that $\sigma_b$ and $W_b$ per broken bond are not constant but decrease with increasing $\xi$. This result explains the superiority of small $f$ networks, for which $\xi$ is smaller and $\sigma_b/\varphi_{bb}$ and $W_b/\varphi_{bb}$ are larger than the networks with large $f$. The data in Figure 7 can be expressed as $\sigma_b/\varphi_{bb} = 117\xi^{-0.2}$ and $W_b/\varphi_{bb} = 52.5\xi^{-0.25}$, although we have no explanation at present.

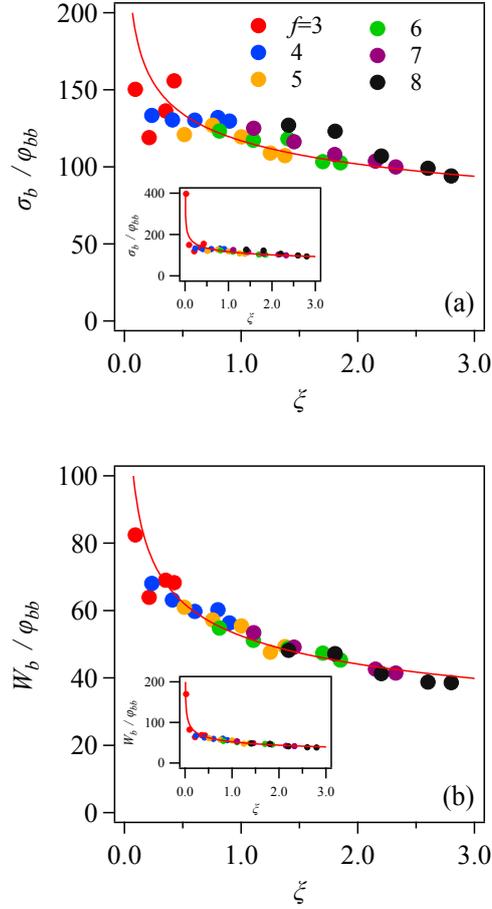

**Figure 7** Stress at break $\sigma_b$ (a) and work for fracture $W_b$ (b) normalized by broken bond fraction $\varphi_{bb}$ plotted against $\xi$. Insets show the plots with the vertical axis range accommodating the single datum for the case with $f = 3$ and $\varphi_c = 0.6$. Red curves display apparent relations written as $\sigma_b/\varphi_{bb} = 117\xi^{-0.2}$ for panel (a) and $W_b/\varphi_{bb} = 52.5\xi^{-0.25}$ for panel (b).

One question that comes up is why small $\xi$ is considered mechanically favorable. One possible explanation is an increase of prepolymers with only two arms reacted with decreasing $\xi$. These prepolymers are excluded in the Scanlan-Case criterion[34,35] from mechanically-effective nodes and strands. Instead, they extend the strand length between effective nodes. Figure 8 (a) shows the number-average strand molecular weight $M_n$ plotted against $\xi$. Here, $M_n$ is normalized by the value for the perfect network $M_0$, i.e., the span molecular weight of the single prepolymer in this study. $M_n/M_0$ data for the examined networks follow a single curve exhibiting that $M_n/M_0$ monotonically decreases with increasing $\xi$, being consistent with the strand length distribution as shown in Appendix B. This result suggests that the fracture characteristics can be described by $M_n/M_0$. We exhibit $\varepsilon_b$ and $W_b/\varphi_{bb}$ plotted against $M_n/M_0$ in Figs. 8 (b) and (c). Consistent with the earlier study[22], $\varepsilon_b$ and $W_b/\varphi_{bb}$ increase with increasing $M_n/M_0$. However, in the range $\xi > 1$, $M_n/M_0$ is almost unity, whereas the fracture characteristics continuously decrease with increasing $\xi > 1$. Thus, the data for large $f$ networks are concentrated around $M_n/M_0 \sim 1$, and further analysis is difficult.

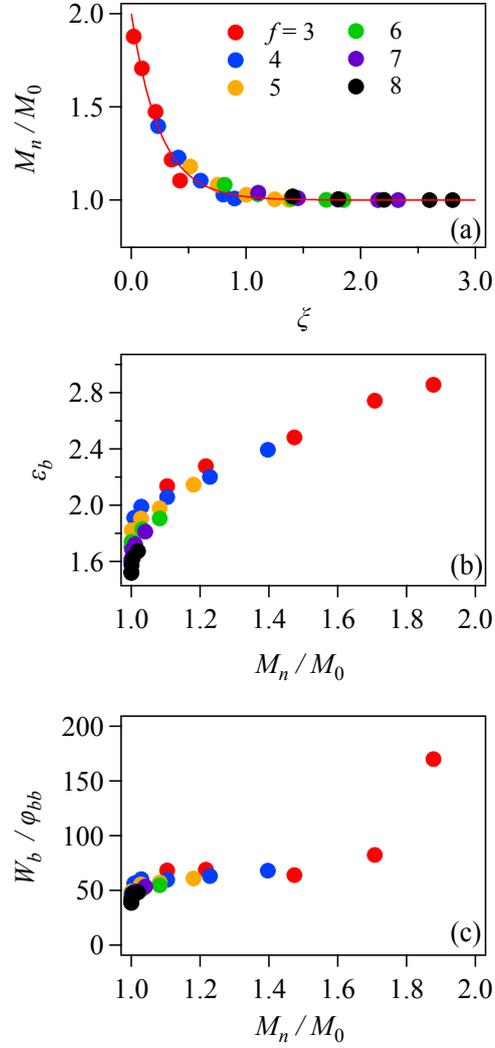

**Figure 8** Panel (a): Number averaged molecular weight of strands between effective network nodes $M_n$ normalized by the span molecular weight of the prepolymer $M_0$ plotted against the cycle rank $\xi$. The red curve shows $M_n/M_0 = 1 + \exp(-4\xi)$. Panel (b) and (c): $\varepsilon_b$ and $W_b/\varphi_{bb}$ plotted against $M_n/M_0$.

**CONCLUSIONS**

We performed a series of phantom chain simulations to investigate the effect of node functionality $f$ on the fracture of polymer networks. We prepared the networks from equimolar binary mixtures of star-shaped prepolymers, for which an end-linking reaction occurs between different chemistries. For simplicity, we employed phantom chains without any interactions between segments. Besides, we eliminated loop formation up to the 3rd-order loops, mainly owing to the binary nature of prepolymers. We simulated equilibration and gelation by the Brownian dynamics scheme to acquire network structures with various $f$ and conversion ratio $\varphi_c$. We imposed energy minimization on the obtained networks and stretched them until the break. We recorded the stress-strain relationship during the stretch and evaluated stress and strain at break and work for fracture, $\varepsilon_b$, $\sigma_b$, and $W_b$. These fracture characteristics strongly depend on $f$ and $\varphi_c$. For the networks with small $f$ values, with increasing $\varphi_c$, $\varepsilon_b$ decreases, whereas $\sigma_b$ and $W_b$ increase steeply. As $f$ increases, the changes in $\varepsilon_b$, $\sigma_b$, and $W_b$ against $\varphi_c$ become mild. Consequently, networks with large $f$ are mechanically superior to networks with small $f$ at small $\varphi_c$, whereas the relation is reversed at large $\varphi_c$. This result explains the contradiction among earlier studies for the effect of $f$ on network fracture. To investigate the origin of the mechanical superiority of small $f$ networks, we measured the cycle rank $\xi$ and the fraction of broken segments $\varphi_{bb}$. The fracture characteristics, $\varepsilon_b$, $\sigma_b/\varphi_{bb}$, $W_b/\varphi_{bb}$ monotonically decrease with increasing $\xi$, exhibiting master curves among the data taken for various $f$ and $\varphi_c$. These results explain the superiority of networks with small $f$, as they have smaller $\xi$ values that give larger $\varepsilon_b$, $\sigma_b/\varphi_{bb}$, and $W_b/\varphi_{bb}$ than networks with large $f$. For a possible explanation, we suggest the role of strand-extending prepolymers that have only two arms reacted.

**ACKNOWLEDGEMENTS**

This study was partly supported by JST-CREST (JPMJCR1992) and JSPS KAKENHI Grant Number 22H01189.

**Appendix A:** Modulus

We obtained modulus from Mooney plots of stress-strain relations, as shown in Figure 9 (a). The Mooney stress exhibits a steady value at a small stretch (i.e., $1/\lambda \sim 1$), followed by an increase with decreasing $1/\lambda$ due to the non-linear spring and rapidly declines to zero due to the network rapture. We obtained the modulus from the stress at the small stretch, as shown by the horizontal broken line, and plotted it against $\varphi_c$ (panel b) and $f$ (panel c). Apart from the non-linearity, the modulus corresponds to the phantom modulus discussed by Lang and Müller[30]. Nevertheless, $G$ increases with increasing $\varphi_c$ and $f$.

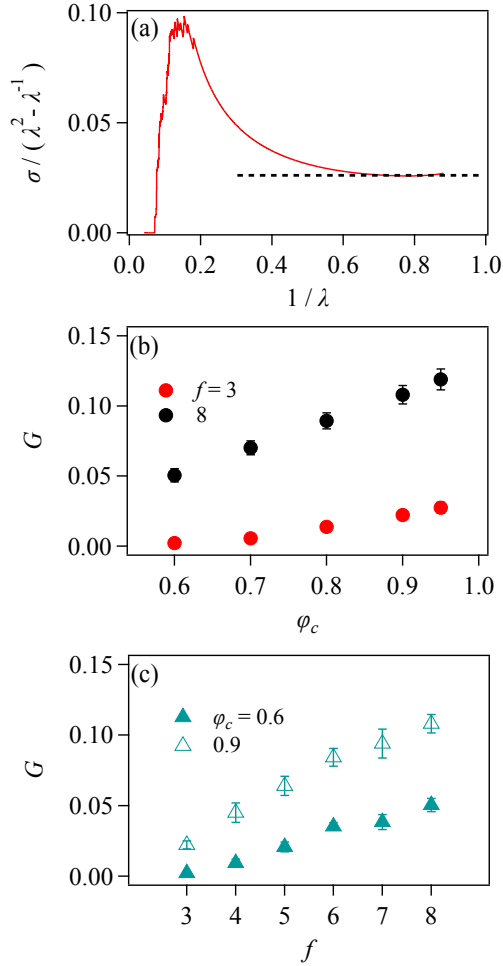

**Figure 9** Panel (a): an example of stress-strain relation in a Mooney plot for $f = 3$, $\varphi_c = 0.9$, and $\rho = 8$. The horizontal broken line indicates the initial modulus $G$. Panel (b): $G$ plotted against $\varphi_c$ for $f = 3$ (red circle) and 8 (black circle). Panel (c): $G$ plotted against $f$ for $\varphi_c = 0.6$ (filled triangle) and 0.9 (unfilled triangle). Error bars in panels (b) and (c) indicate the standard deviation among eight simulation runs.

The behavior of $G$ can be explained by the cycle rank $\xi$, as shown in Fig. 10. If we normalize $G$ by the number density of prepolymers $v_p$, we can nicely superpose the data as shown in panel (b). We observe a non-linearity in the large $\xi$ regime because energy-minimized networks with large $f$ and $\varphi_c$ values exhibit strand stretch (even before macroscopic stretch), as Appendix B shows.

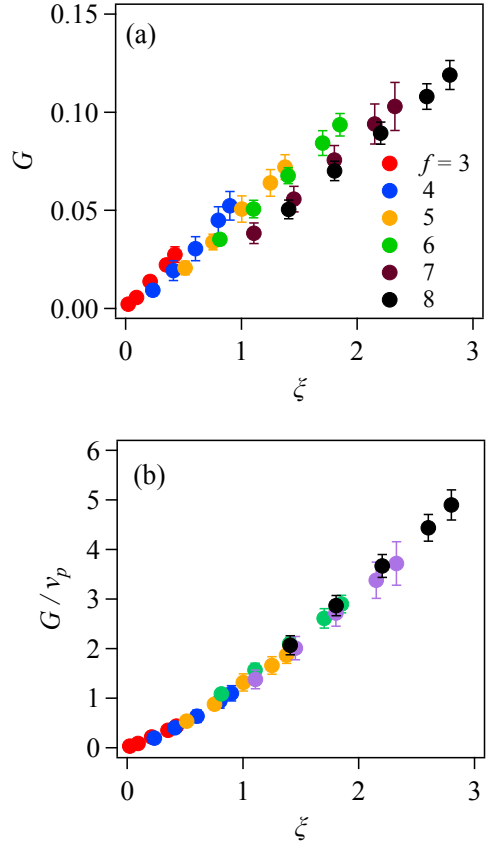

**Figure 10** Mooney modulus $G$ (a) and that normalized by prepolymer density $v_p$ (b) for various $f$ and $\varphi_c$ plotted against the cycle rank $\xi$. Error bars indicate the standard deviation among eight simulation runs.

The results in the main text imply that fracture properties are related to modulus, as all of those quantities are described by $\xi$. One may argue that this argument contradicts earlier studies in which $f = 3$ networks shows superior fracture characteristics than $f = 4$, even if the modulus is the same[11,18]. However, we note that the master curves are observed for $\sigma_b/\varphi_{bb}$ and $W_b/\varphi_{bb}$, not for $\sigma_b$ and $W_b$. Besides, the master curve shown in Fig 10 (b) is for $G/v_p$, and the plot becomes somewhat scattered if we plot $G$ against $\xi$ as shown in Fig 10 (a). Figure 11 shows the stress-strain curves for the networks where we attain similar $G$ values by changing $\varphi_c$. Consistent with the earlier studies[11,18], the comparison demonstrates that $\varepsilon_b$ (and thus $\sigma_b$, and $F_b$) decreases with increasing $f$.

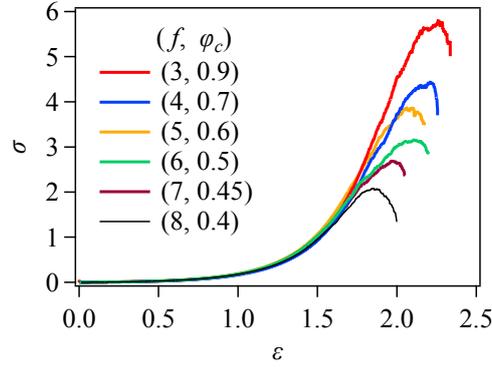

**Figure 11** Stress-strain curves for various $f$ and $\varphi_c$ that realizes similar initial slope to $f=3$ and $\varphi_c=0.9$. The employed $f$ and $\varphi_c$ values are indicated in the figure. Each curve is an ensemble average among eight independent simulation runs.

**Appendix B:** Strand length distributions

As mentioned in the introduction, network stretchability has been discussed concerning the ratio of the maximum stretch to the initial length of strands embedded in the network. To check this argument, in Fig 12, we plot the strand length distribution between branch points $P(r)$ with $\varphi_c=0.9$. For the energy-minimized networks for which solid curves show the distributions, the strand length systematically increases with increasing $f$ due to the increase of $M_n/M_0$. Since the maximum stretch for strand scission is the same for all the cases, this $f$-dependent strand stretch embedded before elongation thus qualitatively explains the $f$-dependence of $\varepsilon_b$ seen in Fig 4. Note that even though $P(r)$ is calculated for all the branch points, not only the effective ones, the contribution of inert branch points and dangling segments is removed by energy minimization. Indeed, when Brownian motion is turned on, due to the inert segments, $P(r)$ becomes insensitive to $f$, as broken curves show.

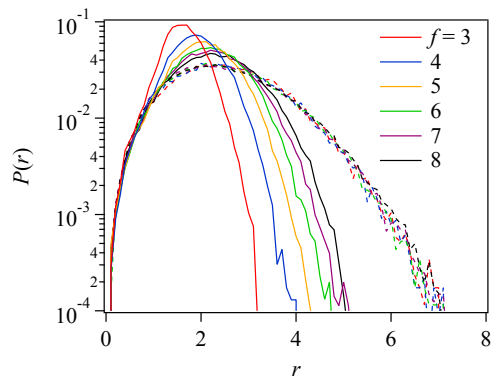

**Figure 12** Strand length distribution for networks after gelation (broken curves) and energy minimization (solid curves) for different $f$ at $\varphi_c=0.9$.